\documentclass[fleqn,useAMS,usenatbib]{mnras}

\usepackage{graphicx}	
\usepackage{amsmath}	
\usepackage{amssymb}	

\usepackage[T1]{fontenc}
\usepackage{ae,aecompl}

\newcommand       \be		{\begin{equation}}
\newcommand       \ee		{\end{equation}}

\title[Optically Thin Core Accretion]{Optically Thin Core Accretion: \\How Planets Get Their Gas in Nearly Gas-Free Discs}
\author[Lee, Chiang, \& Ferguson]{Eve J. Lee$^{1,2}$\thanks{Contact e-mail: \href{mailto:evelee@caltech.edu}{evelee@caltech.edu}}, Eugene Chiang$^{1,3}$, Jason W. Ferguson$^{4}$\\
$^{1}$Astronomy Department, University of California Berkeley, Berkeley, CA 94720-3411, USA\\
$^{2}$TAPIR, Walter Burke Institute for Theoretical Physics, Mail Code 350-17, Caltech, Pasadena, CA 91125, USA\\
$^{3}$Department of Earth and Planetary Science, University of California Berkeley, Berkeley, CA 94720-4767, USA\\
$^{4}$Physics Department, Wichita State University, Wichita, KS 67260-0032, USA
}

\date{}
\pagerange{\pageref{firstpage}--\pageref{lastpage}} \pubyear{2018}
\begin{document}
\maketitle
\label{firstpage}

\begin{abstract}
Models of core accretion assume that in the radiative zones
of accreting gas envelopes, radiation diffuses.
But super-Earths/sub-Neptunes (1--4$R_\oplus$, 2--20$M_\oplus$)
point to formation conditions that are optically thin:
their modest gas masses are accreted from 
short-lived and gas-poor nebulae reminiscent of
the transparent cavities of transitional discs.
Planetary atmospheres born in such environments can be optically thin
to both incident starlight and internally generated
thermal radiation.
We construct time-dependent models of such atmospheres, 
showing that super-Earths/sub-Neptunes can accrete their $\sim$1\%-by-mass gas
envelopes, and super-puffs/sub-Saturns their $\sim$20\%-by-mass
envelopes, over a wide range of nebular depletion histories
requiring no fine tuning.
Although nascent atmospheres can exhibit stratospheric temperature inversions
effected by atomic Fe and various oxides that absorb strongly
at visible wavelengths, the rate of gas accretion remains controlled by the
radiative-convective boundary (rcb) at much greater pressures.
For dusty envelopes, the temperature at the rcb
$T_{\rm rcb} \simeq 2500$ K is still set by 
${\rm H}_2$ dissociation; for dust-depleted envelopes, $T_{\rm rcb}$
tracks the temperature of the visible or thermal photosphere,
whichever is deeper, out to at least $\sim$5 AU.
The rate of envelope growth remains largely unchanged between the old
radiative diffusion models and the new optically thin models, 
reinforcing how 
robustly super-Earths form as part of the endgame chapter in disc evolution.
\end{abstract}


\section{Introduction}
\label{sec:introduction}

The {\it Kepler} mission has revealed that about half of all Sun-like stars
harbor at least one planet with an orbital period $< 85$ days
\citep{howard10,batalha13,petigura13,dong13,fressin13,rowe14,burke15}. 
The most common of these are the super-Earths, here defined
as having radii between 1 and 4 $R_\oplus$.\footnote{Planets with radii between $\sim$2--4 $R_\oplus$ are sometimes called ``mini-Neptunes'' to distinguish
them from $\sim$1--2 $R_\oplus$ super-Earths.
These two populations are indeed observed to be distinct \citep[e.g.,][]{fulton17},
a consequence of the former having retained their atmospheres
and the latter have lost them (e.g., \citealt{owen13,owen17,ginzburg17}; see also our section \ref{ssec:accr_depl}).
Since we are interested in how these planets acquired their atmospheres
in the first place (pre-mass loss), we will not make this distinction but lump
all under the common banner of super-Earths.} Transit-timing analyses
\citep{wu13} and radial velocity surveys \citep{weiss14} have determined
that the masses of super-Earths range from $\sim$2 to 
20$M_\oplus$, with most having masses $\lesssim 6 M_\oplus$.
To be consistent with the observed masses and radii, 
the largest-size super-Earths
must have gas envelopes overlaying their solid cores.
Gas-to-core mass ratios (GCRs) are estimated to be between
$\sim$0.1--10\%, and are more typically $\sim$1\% \citep{lopez14,wolfgang15}.

The GCRs of super-Earths are neither as low as those of solar system
terrestrial planets, nor as high as those of gas giants.
They imply a formation history intermediate in time
between these populations:
neither so late that the parent 
circumstellar disc has lost all its gas, nor so early that
the disc is still gas-rich.\footnote{
More specifically, we posit that cores form during the ``giant
impact'' era when gas dynamical friction was sufficiently weak to
permit protocores to cross orbits and merge; relevant nebular densities
are lower than those of the standard solar-composition
minimum-mass disc by four orders of magnitude and possibly more
depending on the exact orbital spacing between protocores
(see, e.g., \citealt{paper3}, their Figures 5 and 6).} 
This scenario, however, would seem to introduce a fine-tuning problem.
Why are GCRs on the order of 1\%?
At what precise point in the parent nebula's depletion history did super-Earths
appear? How can we guarantee that at that point 
in history there was enough disc gas 
for super-Earth cores to acquire their percent-by-weight atmospheres?

This problem of having to fine-tune
the nebular density is illusory.
Recent studies of how planets get their gas
\citep{paper1,paper2,paper3,ginzburg16} have shown that
the rate at which nebular gas accretes onto cores is 
remarkably insensitive to the nebular density. 
According to these
models, the nebula can be depleted
over a wide range of values---anywhere from one to nine
orders of magnitude
relative to a standard solar-composition ``minimum-mass'' disc---and
super-Earths will still emerge with final GCRs
of 1--10\% (e.g., Figures 4--6 of \citealt{paper3}).
Nor is there any particular fine-tuning problem in time:
the duration of gas accretion can last anywhere from
$\sim$0.1--1 Myr, with even shorter durations allowed
for GCR $\sim 0.1$\%. 
These gas accretion timescales fit nicely
with the timescales over which discs
finally clear; i.e., the duration of the ``transitional'' disc
phase, which is about 10\% of the total disc lifetime 
(Figure 9 of \citealt{owen11}; \citealt{alexander14};
see also Figure 11 of \citealt{espaillat14}).

The fundamental reason why percent-by-weight atmospheres
abound is because cooling times of percent-by-weight atmospheres 
are of order gas disc dispersal times of $\sim$0.1--1 Myr.
Planetary atmospheres can only grow as much as they can cool.
The envelope cooling (= accretion) history is insensitive to the
exact value
of the outer nebular density because the rate of cooling is not controlled
near the Hill or Bondi sphere radius where the atmosphere connects to the
nebula; the cooling is controlled
instead much deeper in the planet's envelope, at its radiative-convective
boundary (rcb).
The rcb is the ``lid'' that regulates how much energy is released
from the planet. That energy is concentrated
in the convective interior, which
dominates the mass of the envelope.
Below the rcb, convection can carry an arbitrarily large energy flux outward.
But above the rcb, the outward flux from radiative diffusion can only be 
as large as the local temperature gradient allows. That gradient equals
its maximum possible value---namely the adiabatic value---at the rcb
(by definition; if it were larger, convection would
ensue, and the rcb would be located at higher altitude).
Thus the rcb acts to throttle
the flux emerging from the convective interior.

The properties of the rcb---its temperature $T_{\rm rcb}$ and
density $\rho_{\rm rcb}$, and by extension its opacity---do not much depend 
on $\rho_{\rm out}$, the density of the nebula at large. If the outermost
portion of a planet's envelope contains significant amounts of dust,
the rcb materializes at the
${\rm H}_2$ dissociation front, and so
$T_{\rm rcb} \simeq 2500\,{\rm K}$ \citep{paper1}.
In a dust-free envelope, the outer radiative layer is nearly isothermal
with the nebula, and so $T_{\rm rcb}$ is set by the external nebular 
temperature $T_{\rm out}$. 
In either case, whether the envelope is dusty or dust-free, the density $\rho_{\rm rcb}$ is controlled essentially by the adiabat in the convective zone below the rcb. Because energy is spent dissociating H$_2$ in the convective zone rather than
heating the gas, the adiabatic index 
$\gamma_{\rm ad}$ is driven closer to unity;  
in particular, $\gamma_{\rm ad} < 4/3$, 
which renders the convective zone centrally concentrated, with the bulk
of the envelope mass residing just above the surface of the underlying
core \citep{paper2}. 
The consequence is that $\rho_{\rm rcb}$ is determined by the envelope
mass, the core mass and radius, the adiabatic index, and $T_{\rm rcb}$ (equation 11 of \citealt{paper2}).
None of these factors depends on $\rho_{\rm out}$.

\begin{figure}
    \centering
    \includegraphics[width=0.5\textwidth]{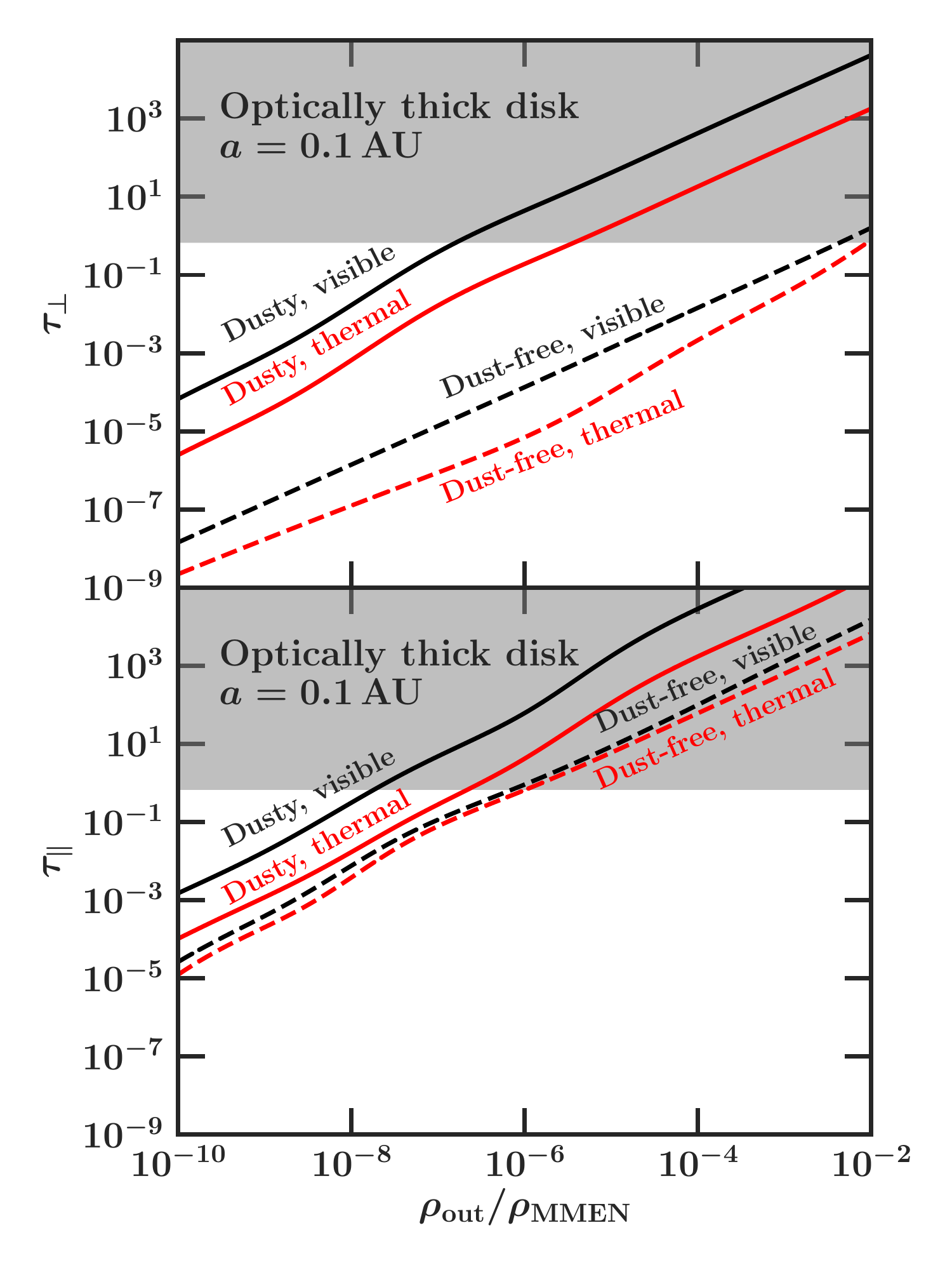}
    \caption{Parent disc densities 
    that are low enough to be relevant for
    the formation of super-Earth atmospheres \citep{paper1,paper3}
    may also be
    low enough to render the disc optically thin.
    Vertical optical depths $\tau_\perp$ are computed for an isothermal disc
    at temperature 1000 K and
    an orbital radius of $a = 0.1$ AU, starting from the midplane and integrating across three vertical scale heights.
    Midplane densities $\rho_{\rm out}$ are normalized to that of the minimum-mass extrasolar nebula, $\rho_{\rm MMEN} = 6\times 10^{-6}\,{\rm g\,cm^{-3}}\,(a/{\rm 0.1 \,AU})^{-2.9}$ \citep{chiang13}.
    Radial optical depths $\tau_\parallel$ are computed by integrating from $a = 0.02$ AU to 0.1 AU.
    Opacities from \citet{ferguson05} are used, assuming metals
    either take the form of grains with an ISM-like size distribution (``dusty'') or are fully in the gas phase (``dust-free'').
    Optical depths at stellar wavelengths
    (``visible'') are evaluated as Planck means at a host
    star temperature $T_\ast = 5800$ K; at wavelengths characterizing disc-generated radiation (``thermal''), they are computed as Planck means at 1000 K. The results of this figure inform our choices for
    $\rho_{\rm out}$ in the rest of this paper; for dusty core accretion
    models, we study $\rho_{\rm out} \leq 10^{-5} \rho_{\rm MMEN}$, and
    for dust-free models, $\rho_{\rm out} \leq 10^{-3} \rho_{\rm MMEN}$ 
    (these choices are motivated by $\tau_\perp$ rather than $\tau_\parallel$
    to explore a wider range of $\rho_{\rm out}$,
    but $\tau_\parallel$ is the more
    relevant for deciding whether planets are exposed directly to stellar
    irradiation).}
    \label{fig1}
\end{figure}

The above analysis would be complete were it not for the
fact that it relies on models that
may occasionally be
unphysical in the following sense. 
Every model of core accretion from the nebula
of which
we are aware (going back to, e.g., \citealt{pollack96}) utilizes
the equation of radiative diffusion to describe energy transport
in the radiative zone of the planetary envelope. The diffusion approximation
is valid wherever the gas is optically thick (in a Rosseland mean sense)
to the local thermal radiation field. Applicable settings include
nascent Jupiters embedded in gas-rich nebulae. 
But super-Earths take us into a new parameter space: they are spawned
from discs that are heavily (but not completely) depleted 
of gas and dust, reminiscent of the inner cavities of transitional discs.
In particular, core formation by giant impacts 
demands nebular gas to be depleted by factors of $\gtrsim 10^4$ to defeat 
gas dynamical friction \citep[see, e.g.,][their Figures 5 and 6]{paper3}.
Not only are such discs potentially optically thin (see Figure \ref{fig1}),
but the planetary envelopes themselves may be as well, both to
incoming starlight and outgoing thermal radiation.

Under optically thin conditions, the diffusion approximation must
be replaced by more sophisticated treatments
of radiative transfer (i.e., treatments more appropriate
for stellar/planetary atmospheres than stellar/planetary interiors).
How does accounting for the outermost, optically thin
layers of planetary envelopes, naked before their host stars,
affect how they grow? How does core accretion play
out in the transparent cavities of transitional discs?
Does the rcb enjoy the same immunity to the outer nebular density
when the layers above it are optically thin?
This paper seeks to answer these questions
by grafting an Eddington two-stream atmosphere \citep{guillot10}
onto our previous core accretion model \citep{paper1,piso14}.
The model is detailed in section \ref{sec:model} and its
output reviewed in section \ref{sec:result}.
Limitations to our opacity tables restrict these main results
to stellocentric distances of $\sim$0.1 AU, where nebular temperatures
are on the order of 1000 K. Nevertheless, in section \ref{ssec:1au},
we experiment with alternative opacities and outline some
semi-quantitative considerations that will enable us to explore
how planets get their gas at larger distances, with particular
attention paid to whether ``super-puff''
planets still need to acquire their gas beyond $\sim$1 AU \citep{paper3}.
We place our work on atmospheric accretion into context with others'
work on atmospheric erosion in section \ref{ssec:accr_depl},
and wrap up in section \ref{sec:conclusions}.

\section{Model}
\label{sec:model}

We study how a rocky core accretes a gas envelope from a disc
that is optically thin to incoming starlight.
Our procedure follows that of \citet{paper1} (see also \citealt{piso14}).
We first construct one-dimensional hydrostatic snapshots of the
planet, each having a unique gas-to-core mass ratio.
Then we string the snapshots together along a timeline
by calculating the time it takes to cool from one snapshot to the next.
What is different in this paper is that we no longer assume the entire
envelope to be optically thick; instead, we account
for how the uppermost regions of the envelope can be
optically thin to incoming visible and/or outgoing thermal
radiation. The goal is to quantify how much the properties of the outer
envelope affect the gas accretion rate.

\subsection{Optically Thin Outer Layer}
\label{ssec:optically_thin_outer_layer}

In the outermost portions of the gas envelope, where material
is optically thin to either incident or internally generated
radiation, we solve the following stellar structure equations:
\begin{equation}
    \frac{dM}{dr} = 4\pi r^2 \rho
    \label{eq1}
\end{equation}
\begin{equation}
    \frac{dP}{dr} = -\frac{GM(<r)}{r^2} \rho
    \label{eq2}
\end{equation}
\begin{equation}
     \frac{d\tau}{dr} = -\kappa_{\rm th}\rho
     \label{eq3}
\end{equation}
\begin{align}
    \frac{4T^3dT}{dr} &= \left[\frac{3}{4}T^4_{\rm int} - \frac{3\sqrt{3}\gamma}{4}T^4_{\rm eq}\left(\frac{\gamma}{\sqrt{3}} - \frac{1}{\gamma\sqrt{3}}\right)e^{-\gamma\sqrt{3}\tau}\right]\frac{d\tau}{dr}
\label{eq4}
\end{align}
where $G$ is the gravitational constant, $r$ is the radius from the center
of the planet, $\rho$, $P$, and $T$ are the density, pressure, and
temperature of the gas, $M(<r)$ is the mass enclosed within $r$,
$\kappa_{\rm th}$ is the opacity to internal thermal emission
(evaluated as a Planck mean; see equation \ref{eq:kap_th} and
surrounding discussion),
$\tau$ is the corresponding radial optical depth (with $\tau=0$
at the outermost radius $R_{\rm out}$, defined as the Hill or Bondi radius, whichever is smaller, and increasing inward to the core).
We will elaborate on $\kappa_{\rm th} (r)$,
and on the constants $T_{\rm int}$, $T_{\rm eq}$, and $\gamma$,
in what follows.

Equation (\ref{eq4}) is derived from the temperature
profile of an Eddington two-stream
atmosphere (e.g., \citealt{guillot10}):
\begin{align}
    T^4 &= \frac{3T^4_{\rm int}}{4}\left[\frac{2}{3} + \tau\right] \nonumber \\
    &+ \frac{3T_{\rm eq}^4}{4}\left[\frac{2}{3}+\frac{1}{\gamma\sqrt{3}}+\left(\frac{\gamma}{\sqrt{3}}-\frac{1}{\gamma\sqrt{3}}\right)e^{-\gamma\sqrt{3}\tau}\right] \,.
\label{eq5}
\end{align}
The atmosphere is heated from below by outgoing internal radiation
(the first term) and from above by incoming stellar radiation
(the second term). 
The internal radiation is generated by cooling of the entire
envelope. The outgoing flux $\sigma T_{\rm int}^4$ is emitted
at the thermal
photosphere $R_{\rm ph,th}$ such that
\begin{equation}
    T_{\rm int}^4 \equiv \frac{L}{4\pi\sigma R_{\rm ph,th}^2}
    \label{eq6}
\end{equation}
where $\sigma$ is the Stefan-Boltzmann constant
and $L$ is the total internally generated luminosity from the envelope.
The quantities $R_{\rm ph,th}$ and $L$ are solved
as part of our numerical procedure; they are not simple inputs.
The incoming stellar flux is averaged over the entire planetary surface:
\begin{equation}
    T_{\rm eq}^4 \equiv \frac{T_{\star}^4}{4}\left(\frac{R_\star}{a}\right)^2
    \label{eq8}
\end{equation}
where $T_{\star}$ is the stellar effective temperature ($=5800$ K
for this paper), $R_\star$ is the stellar radius ($=1 R_\odot$),
and $a$ is the planet's orbital distance ($=0.1$ AU).
For these parameter choices, $T_{\rm eq} = 883$ K.

Whether the incoming stellar or the outgoing internal flux contributes
more to the heating at a given location depends on the ratio of the
``visible'' and ``thermal'' opacities:
\begin{equation}
    \gamma \equiv \frac{\kappa_{\rm vs}}{\kappa_{\rm th}}
    \label{eq7}
\end{equation}
where $\kappa_{\rm vs}$, evaluated
as a Planck mean (see equation \ref{eq:kap_vs}
and surrounding discussion), is the opacity to 
the bulk of the radiation coming from the star 
(which is at visible wavelengths for our choice of $T_\ast = 5800$ K,
but can in principle be in any wavelength range).
When the envelope has strong absorbers in the visible
($\gamma \gg 1$), the stellar flux tends to
dominate the heating
($T^4 \sim T_{\rm eq}^4 \gamma e^{-\gamma \sqrt{3}\tau}$; temperature decreases inward).
In the opposite limit ($\gamma \ll 1$), the internal flux
tends to control the heating ($T^4 \sim T_{\rm int}^4(2/3+\tau)$;
temperature increases inward).

Equation (\ref{eq5}) assumes that $\gamma$ is spatially constant;
we choose to evaluate it at the outermost radius $R_{\rm out}$.
Although $\gamma$ is held constant in a given
snapshot, $\kappa_{\rm th}$ is calculated anew for every position $r$;
by extension, $\kappa_{\rm vs}$ also varies with depth.
We now turn to how we compute $\kappa_{\rm th}$ and $\kappa_{\rm vs}$
(the latter only at $R_{\rm out}$).

\begin{figure}
    \centering
    \includegraphics[width=0.5\textwidth]{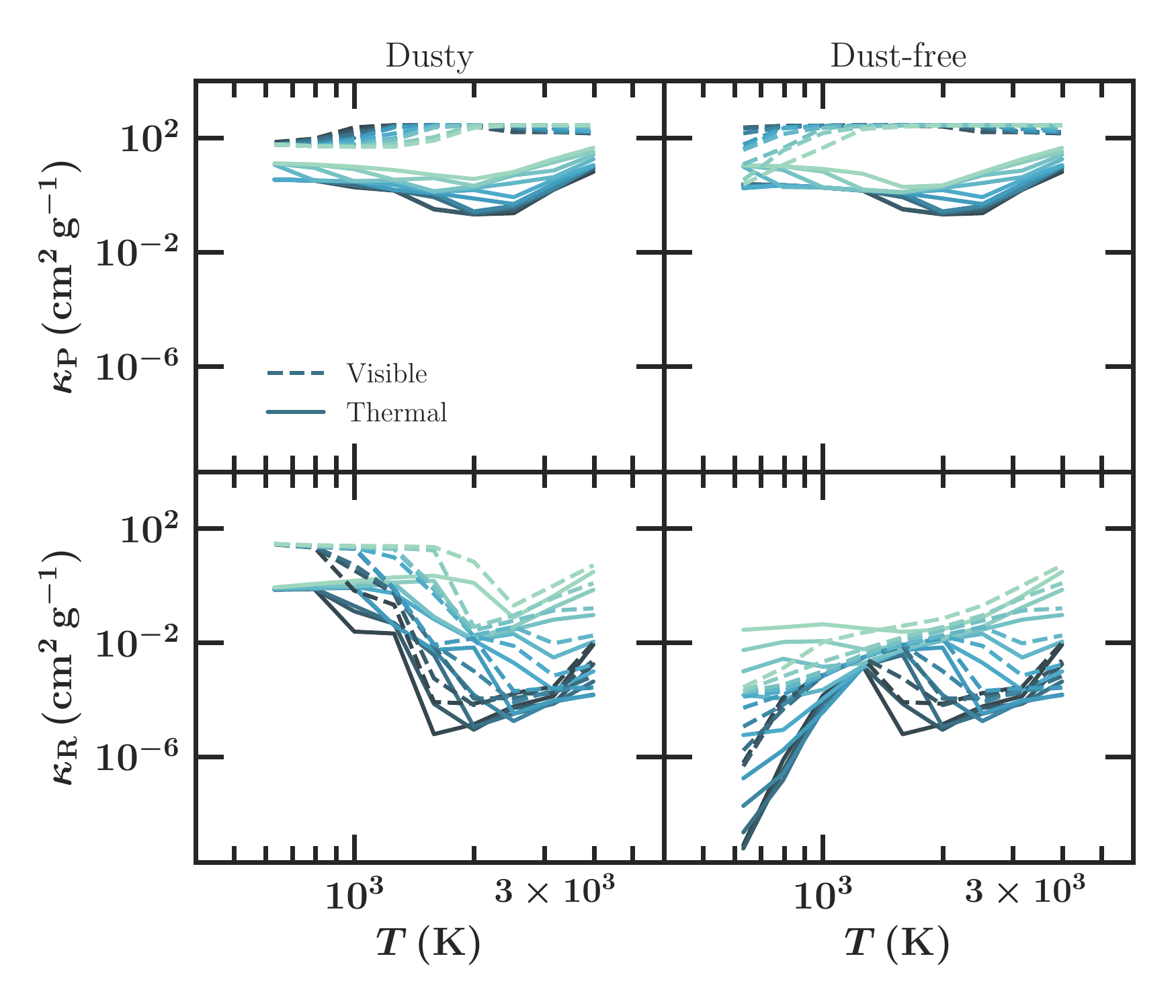}
    \caption{Opacities vs.~temperature and density. 
    Dark blue to light green colors correspond
    to $\log \rho \,({\rm g\,cm^{-3}}) \in [-17, -15.5, -14, \dots, -3.5]$.
    As $T$ varies on the abscissa, the equilibrium chemical composition 
    of the gas (+ dust in dusty models) changes according to the equilibrium 
    calculations of \citet{ferguson05}. Opacities are computed as either
    Planck ($\kappa_{\rm P}$) or Rosseland ($\kappa_{\rm R}$) averages;
    ``visible'' opacities evaluate the blackbody radiation field
    at the stellar temperature $T_\ast = 5800$ K, while ``thermal'' opacities
    evaluate the blackbody function at the local $T$.
    Only Planck means (shown in the top two panels) are used for our
    two-stream, optically thin outer layer (sections
    \ref{ssec:optically_thin_outer_layer} and
    \ref{ssec:opacity}); we revert to Rosseland means below the visible-light
    or thermal photosphere, whichever is deeper.
    Over most temperatures and densities, the visible $\kappa_{\rm P}$
    is two orders of magnitude greater than the thermal $\kappa_{\rm P}$
    (hence $\gamma \equiv \kappa_{\rm vs}/\kappa_{\rm th} \sim 240$,
    as indicated in Figure \ref{fig5}), with
    dust making little difference since the Planck mean is dominated by 
    absorption lines of gaseous atoms and molecules (Figure \ref{fig3}).}
    \label{fig2}
\end{figure}

\begin{figure*}
    \centering
    \includegraphics[width=\textwidth]{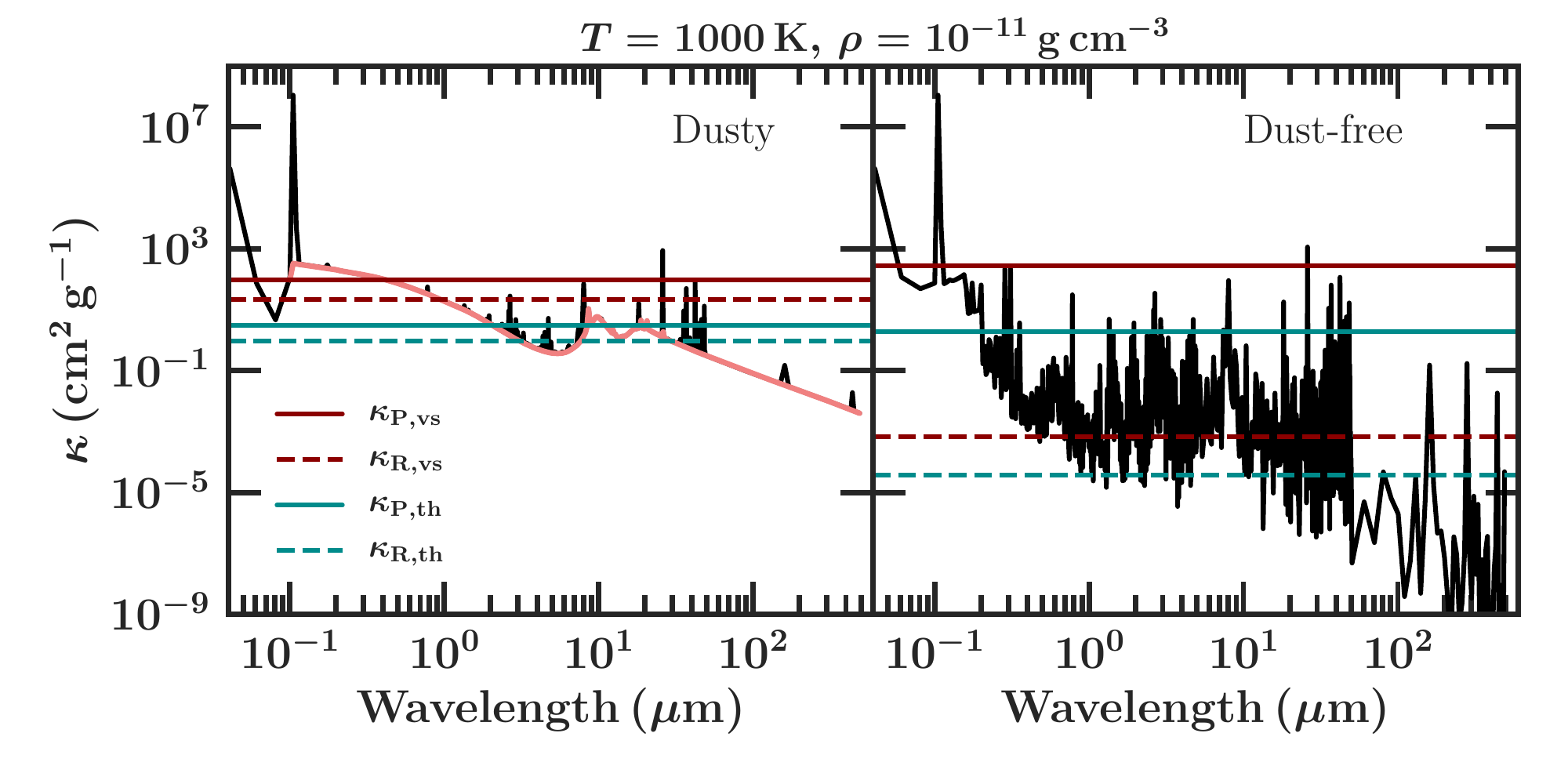}
    \caption{Opacity vs.~wavelength (black curve) at fixed temperature
    $T = 1000\,{\rm K}$ and density $\rho = 10^{-11}\,{\rm g\,cm^{-3}}$.
    Horizontal lines mark mean opacities, evaluated either as
    Planck (P) or Rosseland (R) averages, with the blackbody radiation
    field evaluated for either $T_\ast = 5800$ K (vs) or $T$ (th).
    We employ Planck means for our optically thin, two-stream atmospheres,
    and Rosseland means for the interior below. Dust (left panel) contributes
    a strong continuum opacity (thick orange curve) that is absent from the
    dust-free model (right panel); consequently, the Rosseland means
    (dashed horizontal lines), which are more sensitive to the continuum
    than to absorption peaks, are much higher in dusty than dust-free models.
    By contrast, the Planck means (solid horizontal lines) do not vary
    much between dusty and dust-free models because they are more sensitive
    to absorption peaks, arising here from atomic Fe ($\sim$0.1 $\mu$m)
    and oxides (TiO between 
    0.4 $\mu$m and 1.2 $\mu$m, CO near 4 $\mu$m, and water at longer wavelengths).}
    \label{fig3}
\end{figure*}

\subsection{Opacities}
\label{ssec:opacity}

Wavelength ($\lambda$) specific
opacities $\kappa(\lambda,\rho,T)$ are generated 
from 
a modified version of the stellar atmosphere code \texttt{PHOENIX}
as described in
\citet{ferguson05}, assuming 
solar metallicity ($Z=0.02$, where elemental
abundances are scaled to those in \citealt{GN93}).
Two different flavours of opacity models are explored: ``dusty,''
in which an ISM-like grain size distribution is assumed, and ``dust-free,''
in which metals never take the form of dust and are instead in the
gas phase at their full abundances.
The Ferguson et al.~opacities incorporate heavy atomic metals
such as iron, and in this regard appear more complete
than other opacity tables \citep[e.g.,][]{freedman14}.
At the low pressures and high temperatures 
characterizing our highly distended gas envelopes,
atomic lines at short wavelengths ($\lesssim 0.2 \,\mu$m)
contribute significantly to $\kappa_{\rm vs}$.
We have found that excluding the heavy atoms
decreases $\kappa_{\rm vs}$ by 1--2 orders of magnitude.\footnote{In an extensional calculation in \S\ref{ssec:1au}, we will replace the \citet{ferguson05} opacities
with the \citet{freedman14} opacities. Although the latter are missing the
gas-phase heavy metals, they do extend to lower temperatures,
enabling us to explore models at 1 AU instead of our standard 0.1 AU.}

The opacities $\kappa_{\rm vs}$ and $\kappa_{\rm th}$
are computed as Planck means:
\begin{equation}
    \kappa_{\rm vs} (\rho, T) \equiv \frac{\int \kappa(\lambda, \rho, T) B(\lambda, T_{\star}) d\lambda}{\int B(\lambda, T_{\star}) d\lambda}
    \label{eq:kap_vs}
\end{equation}
\begin{equation}
    \kappa_{\rm th} (\rho, T) \equiv \frac{\int \kappa(\lambda, \rho, T) B(\lambda, T) d\lambda}{\int B(\lambda, T) d\lambda} \,,
    \label{eq:kap_th}
\end{equation}
the former using the Planck function $B$ evaluated at the stellar
temperature $T_\star$, and the latter at the local gas temperature $T$.
The use of Planck means is not entirely justified.
In the derivation by Guillot (\citeyear{guillot10};
see, e.g., his equation 16),
the wavelength-averaged opacities are weighted by the
mean intensity $J$. The problem is that
$J$ is not known {\it a priori}. We have chosen to replace $J$
with the local Planck intensity $B$. An alternative is to 
replace equation (\ref{eq:kap_vs}) and/or (\ref{eq:kap_th}) with
the Rosseland mean \citep[e.g.,][]{rogers11}, but the Planck
mean seems preferable since we are considering regions optically
thin to both incoming starlight and outgoing internal radiation.
\citet{hubeny03} also use Planck mean opacities for their
atmospheres.

In principle, one need not adopt the same kind of mean for both
$\kappa_{\rm vs}$ and $\kappa_{\rm th}$.
For example, if the envelope is 
optically thick to visible light but optically thin to its own thermal
radiation, then it would seem appropriate to use the
Rosseland mean for $\kappa_{\rm vs}$ and the Planck mean
for $\kappa_{\rm th}$. But this choice can lead to
internal inconsistencies; as Figures \ref{fig2} and \ref{fig3} show,
the Rosseland mean $\kappa_{\rm vs}$ can be significantly smaller
than the Planck mean $\kappa_{\rm th}$,
which would not fit with our original supposition
that the atmosphere be thick in the visible
and thin in the infrared.
We avoid this problem and simply adopt Planck means for both
$\kappa_{\rm vs}$ and $\kappa_{\rm th}$. 
Our choice of Planck means also accords 
better with our assumption 
of a spatially constant $\gamma$; in Figure \ref{fig2},
we see that our Planck means change much less than our Rosseland
means with temperature and density. Thus our Planck
means afford our two-stream solution greater self-consistency.

The ambiguities
in the choice of mean opacities underscore the limitations of using
a constant-$\gamma$ atmosphere to describe
what is really more complicated. Although the two-stream
model used here cannot claim accuracy,
it hopefully suffices to explore, in a qualitative way, 
the effect of the outermost
atmospheric layers on the gas accretion rate.
Comparison with full radiative transfer models 
would be welcome.

Figure \ref{fig2} shows $\kappa_{\rm vs}$ and $\kappa_{\rm th}$
for a range of densities and temperatures.
Dusty and dust-free models feature similar Planck mean opacities
(compare top left and top right panels). In particular,
the dusty model betrays no drop in opacity at $T \gtrsim 1500$ K,
despite dust grains sublimating at those temperatures.
Planck means (unlike Rosseland means) are sensitive to opacity
peaks, and not to the continuum opacity controlled by dust.
The strongest peaks are from gas absorption lines: from atomic
iron at $\sim$0.1 $\mu$m and molecular species (e.g., CO, H$_2$O,
and oxides like TiO) at 1--40 $\mu$m wavelength.
The iron lines are also responsible for rendering $\kappa_{\rm vs}$
1--2 orders of magnitude higher than $\kappa_{\rm th}$.
In the literature, TiO and VO (which our opacity tables
include) are often quoted as the 
strongest absorbers in the visible 
\citep[e.g.,][]{hubeny03}, but this claim is based on models
that include only alkali metals, and not iron, in their repository of
atoms. Atmospheres like ours that absorb strongly in the visible and
weakly in the infrared (i.e., atmospheres with $\gamma \gg 1$)
exhibit temperature inversions, as we will see in section \ref{sec:result}.

\subsection{Overall Procedure}
\label{ssec:procedure}
Our procedure is the same as that in \citet{paper1}, except that in addition
to iterating on $L$, we also iterate on $T_{\rm int}$
in the construction of each snapshot. Only a
particular pair of values $L$ and $T_{\rm int}$ can satisfy
equations (1)--(4) for a given set of outer boundary conditions,
a given core mass $M_{\rm core}$,
and a given gas-to-core mass ratio
GCR $\equiv M_{\rm gas}/M_{\rm core}$.
For all models in this paper, we fix $M_{\rm core} = 5\, M_\oplus$,
$R_{\rm core} = 1.6 R_\oplus$,
$a = 0.1$ AU, $T_{\rm eq} = 883$ K,
and a nebular temperature $T_{\rm out} = 1000$ K.
The outer nebular density $\rho_{\rm out}$ is scanned over
a range of values whose maximum is that of a fiducial gas-rich
disc, $\rho_{\rm MMEN} = 6 \times 10^{-6}\,{\rm g\,cm^{-3}}$.
For these parameter choices, following section 2.1.1 of
\citet{paper1}, $R_{\rm out}$ equals the Hill radius of the planet
($\sim$$40\,R_\oplus$).

For the above parameters, and for a given GCR, we proceed as follows:

\begin{enumerate}
\item Take the density of the envelope
``surface'' at $r = R_{\rm out}$ to equal the
nebular density $\rho_{\rm out}$.
\item Guess $L$.
\item Calculate a provisional value for
$R_{\rm ph,th}$ as that radius where the
(thermal, Planck mean) optical depth $\tau = 2/3$,
assuming an isothermal outer atmosphere
at temperature $T = T_{\rm eq}$ and integrating inward
from $R_{\rm out}$. Using this $R_{\rm ph,th}$ and $L$, calculate
a provisional value for $T_{\rm int}$ using (\ref{eq6}).
\item Obtain the envelope surface temperature
(not to be confused with the nebular temperature $T_{\rm out}$)
by inserting $\tau=0$ into (\ref{eq5}).
\item Evaluate $\gamma$ using the envelope
density and temperature at $\tau = 0$ (and Planck mean opacities).
\item Integrate equations (\ref{eq1})--(\ref{eq4}) from $R_{\rm out}$ down
to either the thermal photosphere $R_{\rm ph,th}$ or the visible-light
photosphere $R_{\rm ph,vs}$, whichever is deeper
(for our model parameters, $R_{\rm ph,th}$ is deeper).
Use Planck mean opacities throughout.
\item From the resultant $R_{\rm ph,th}$, re-calculate $T_{\rm int}$
using (\ref{eq6}). If this $T_{\rm int}$ matches the provisional
value of step (iii)
(within 1\%), then proceed to the next step; otherwise,
return to step (iv) using this $T_{\rm int}$.
\item Continue to integrate the stellar structure equations
down to the radius of the solid core, swapping out equations (\ref{eq3})
and (\ref{eq4}) for the usual equations of radiative diffusion
and convective energy transport (equations 5--8 of \citealt{paper1},
with opacities now evaluated as Rosseland means).
\item If the resultant envelope gas mass
matches the desired value to within some
tolerance ($\sim$$10^{-5}$--$10^{-6} M_\oplus$),
then the hydrostatic snapshot is complete;
otherwise, return to step (ii).
\end{enumerate}

Snapshots are constructed in order of increasing GCR
and placed on a timeline that extends up to $t_{\rm disc} = 1$ Myr,
the assumed disc lifetime (more precisely,
the time over which $\rho_{\rm out}$ is held constant,
after which the nebular density is assumed to fall to zero).

\begin{figure}
    \centering
    \includegraphics[width=0.5\textwidth]{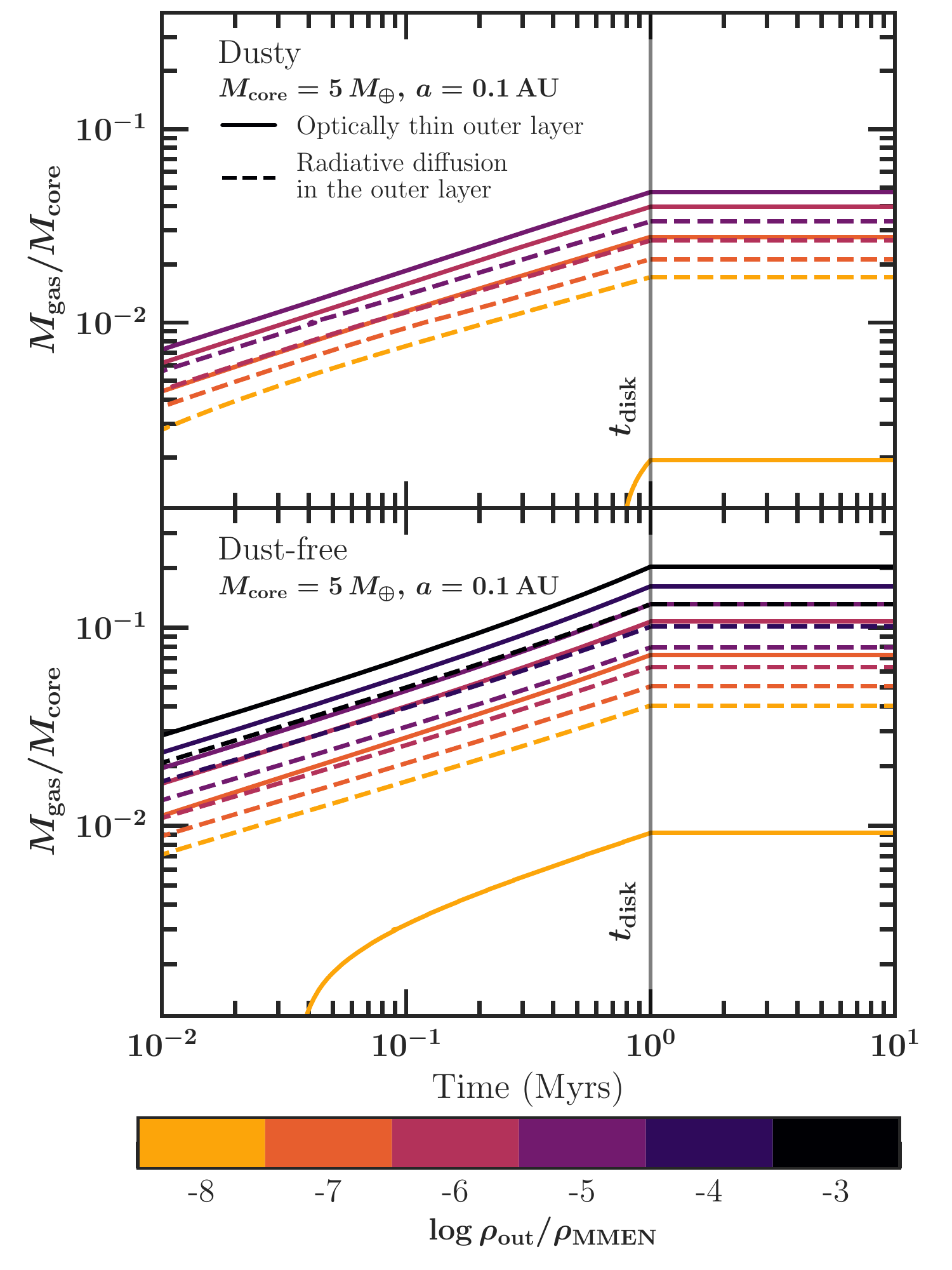}
    \caption{How close-in super-Earths
    ($M_{\rm core} = 5 M_\oplus$ at $a = 0.1$ AU)
    get their gas. As long as the nebular density
    $\rho_{\rm out}$ exceeds $10^{-8} \rho_{\rm MMEN}$ (see color bar),
    our new models
    that treat more carefully the planet's outermost optically thin layers
    (solid curves) do not much differ from our previous models
    that rely only on the diffusion equation to describe radiative layers
    (dashed curves). Final gas-to-core mass ratios of $\sim$1--10\%,
    similar to those inferred from observations,
    result over many orders of magnitude variation in the background
    nebular density. Gas accretion is halted
    at an assumed disc lifetime of $t_{\rm disc} = 1$ Myr;
    it is clear from the displayed curves
    that shifting $t_{\rm disc}$ to 0.1 Myr
    (or even shorter times) would still succeed in generating
    percent-by-weight atmospheres.
    There are fewer curves in the dusty (top) panel
    because dust grains render atmospheres more opaque;
    dusty envelopes develop outer optically thin layers only for
    the more extreme nebular depletion factors (see Figure \ref{fig1}).}
    \label{fig4}
\end{figure}

\begin{figure*}
    \centering
    \includegraphics[width=\textwidth]{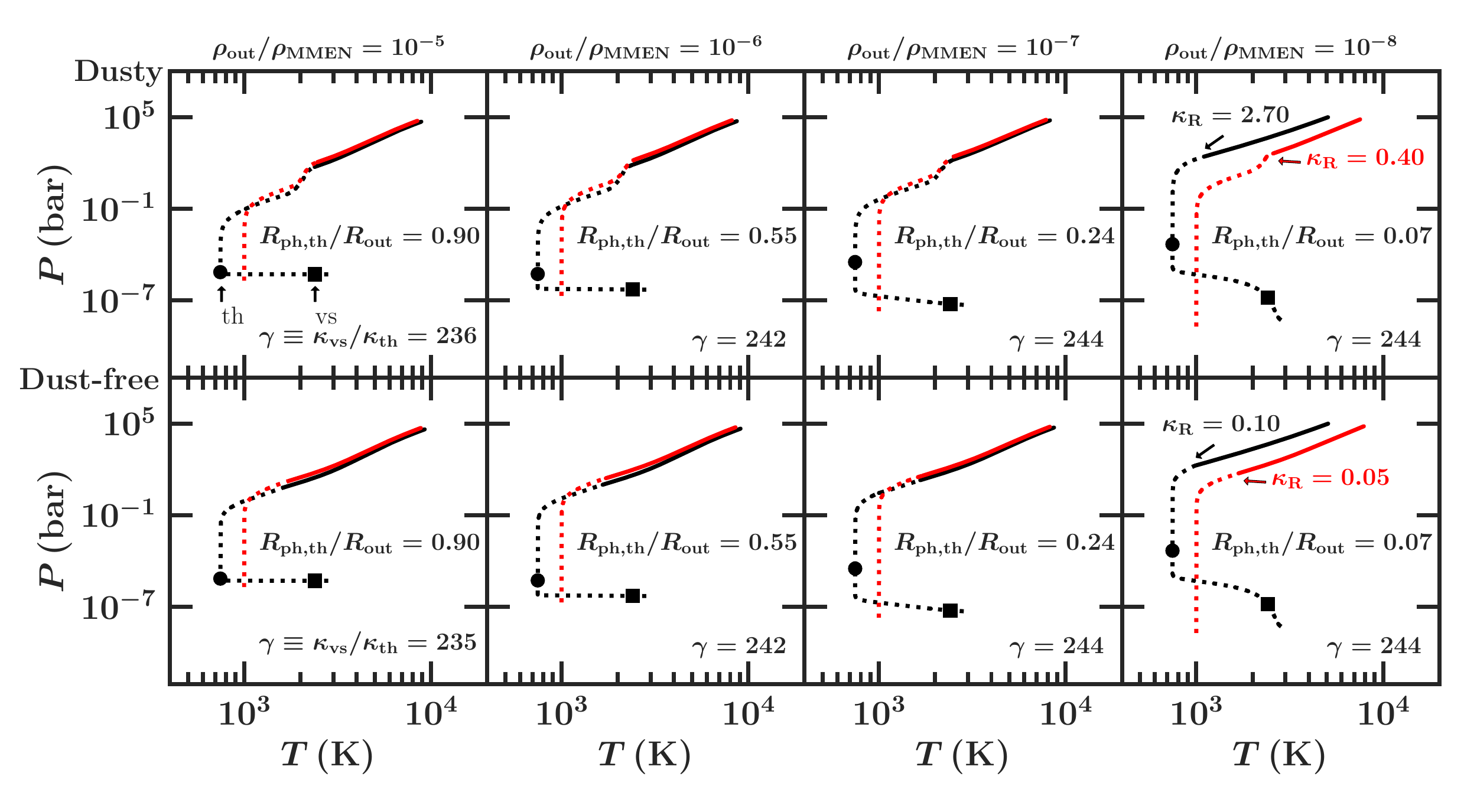}
    \caption{Atmospheric $P$-$T$ profiles of optically thin (black)
    and more primitive radiative diffusion (red) models at fixed
    ${\rm GCR} \equiv M_{\rm gas}/M_{\rm core} = 0.15 M_\oplus / 5 M_\oplus = 0.03$.
    The visible (vs) and thermal (th) photospheres 
    (where $\tau = 2/(3\sqrt{3}\gamma)$ and $2/3$, respectively)
    are marked by squares and
    circles, respectively. Because
    $\gamma = \kappa_{\rm vs}/\kappa_{\rm th} \gg 1$,
    the visible photosphere lies above the thermal, and there
    are strong temperature inversions captured only by the optically thin
    model. Regardless, the radiative-convective boundaries---where the
    radiative zones (dashed lines) connect with the convective zones
    (solid lines)---are practically identical 
    between the optically thin and radiative diffusion models. Over a wide
    range of nebular depletion factors
    $\rho_{\rm out}/\rho_{\rm MMEN} \geq 10^{-7}$,
    pressures and temperatures of the rcb's differ by at most factors of 2
    between the two classes of model. Only when
    $\rho_{\rm out}/\rho_{\rm MMEN} = 10^{-8}$ do the model rcb's diverge
    strongly: the optically thin model features a higher density at its rcb,
    and thus a higher opacity (as annotated in cgs units) and a lower cooling/accretion rate.
    At these ultra-low nebular densities, differences between the two
    radiative transfer treatments become accentuated as the rcb is
    pushed close to the rocky core
    and the envelope becomes nearly completely radiative.}
    \label{fig5}
\end{figure*}

\begin{figure}
    \centering
    \includegraphics[width=0.5\textwidth]{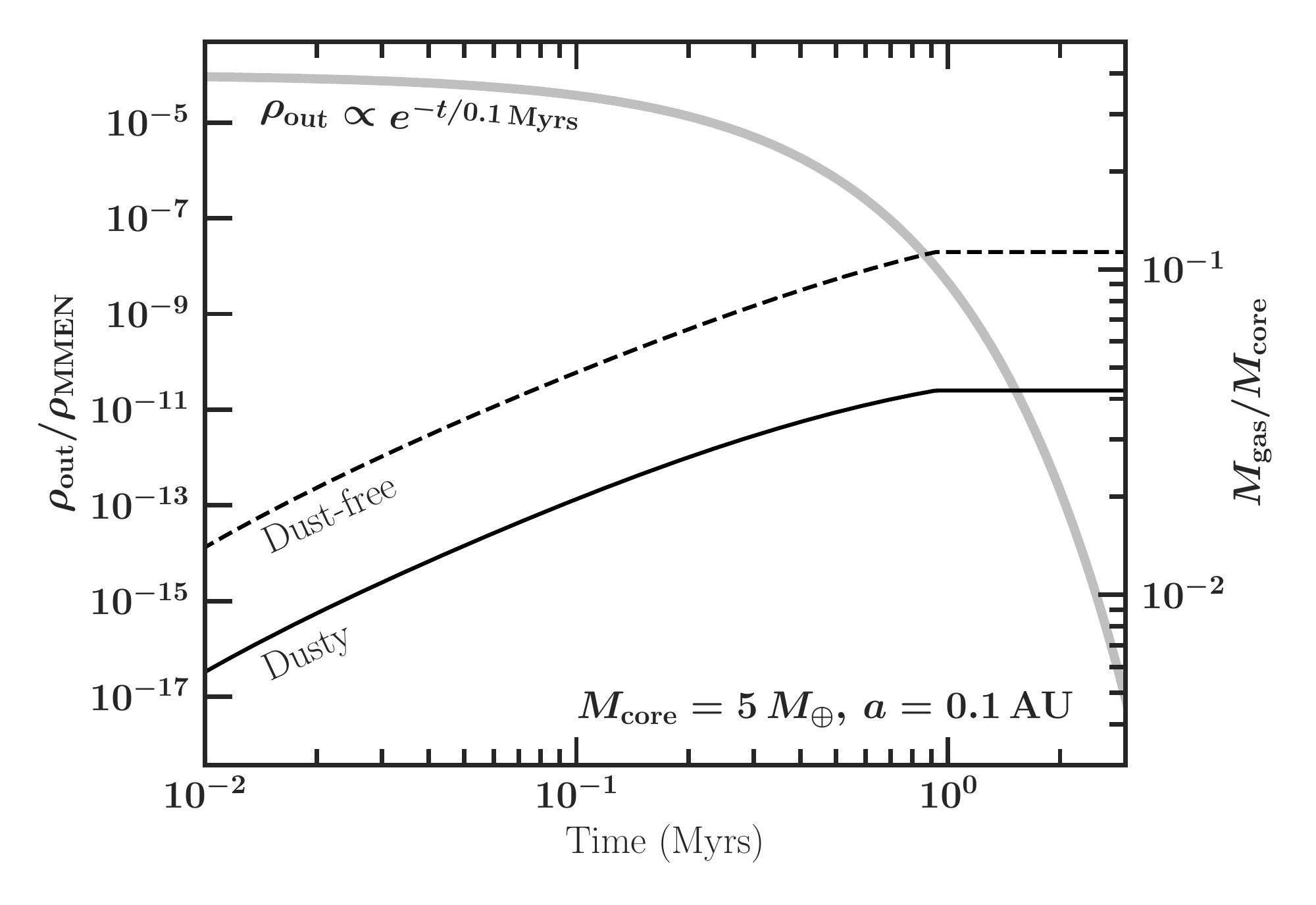}
    \caption{How 5$M_\oplus$ cores build their envelopes in an
        exponentially depleting nebula. The gas-to-core mass ratio
        (GCR $\equiv M_{\rm gas}/M_{\rm core}$; solid and dashed
        curves using the right-hand axis) are computed by first
        fitting a scaling relationship to our numerical results for
        GCR vs.~nebular density (we find GCR $\propto \rho_{\rm out}^{0.1}$),
        and then integrating over the time-derivative of GCR as the
        nebular gas depletes with an e-folding time of 0.1 Myr (gray
        curve using the left-hand axis). 
        We halt envelope growth once 
        $\rho_{\rm out}/\rho_{\rm MMEN}$ falls below $10^{-8}$ since the rate of gas
        accretion slows considerably 
        (by about an order of magnitude compared to $\rho_{\rm
            out}/\rho_{\rm MMEN} = 10^{-7}$; see Figure \ref{fig4})
        at this level of depletion.}
    \label{fig:gcr_v_t_evoldisk}
\end{figure}

\begin{figure}
    \centering
    \includegraphics[width=0.5\textwidth]{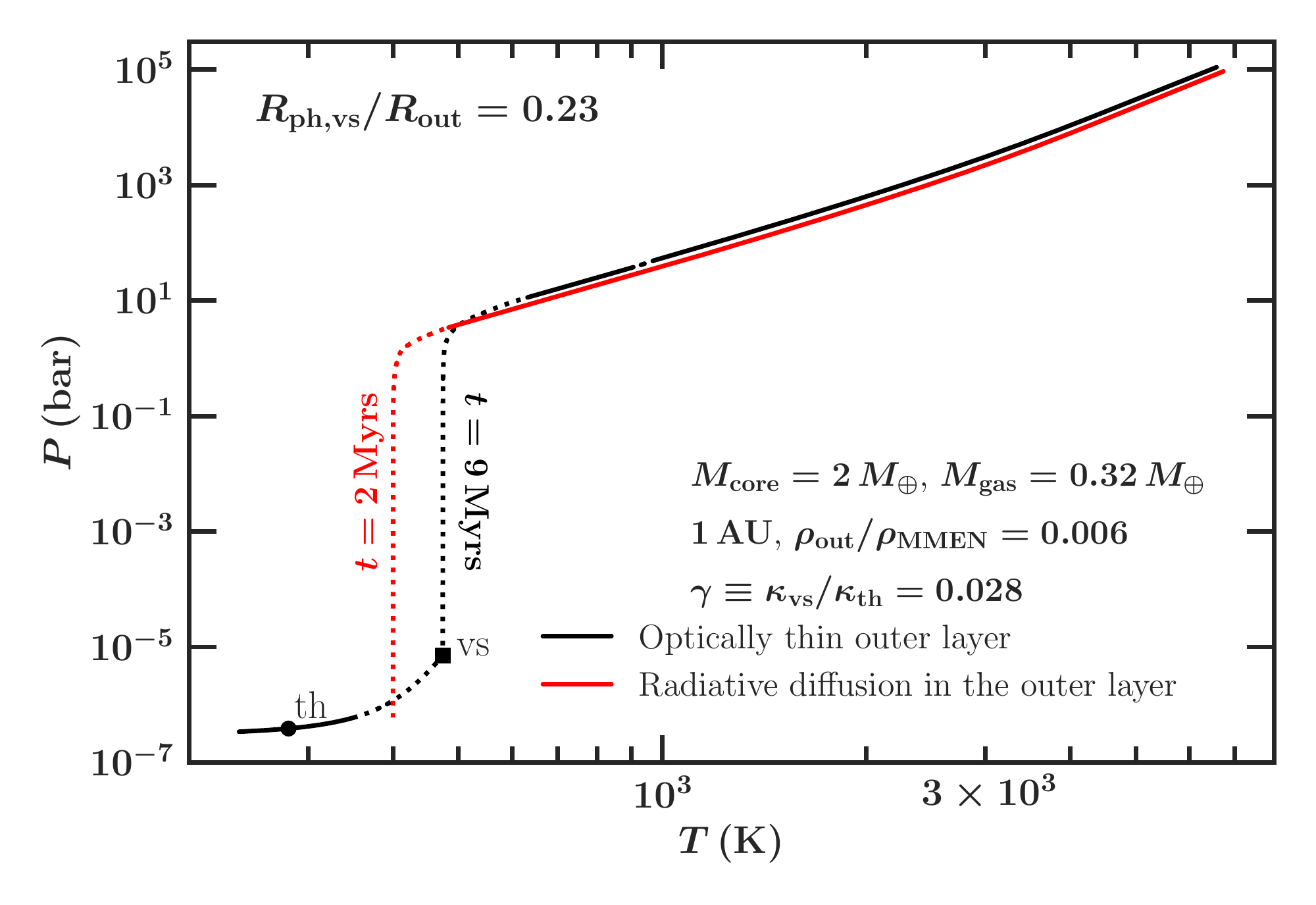}
    \caption{Atmospheric $P$-$T$ profile of a nascent super-puff
    ($2 M_\oplus$ core, GCR = 16\%) in a depleted disc at 1 AU
    ($\rho_{\rm out} = 0.006 \,\rho_{\rm MMEN} =
    4 \times 10^{-11} \, {\rm g\,cm^{-3}}$).
    Figure notation is identical to Figure \ref{fig5}.
    The similarity of the radiative-convective boundaries
    between the optically thin and radiative diffusion
    models implies that the former's more careful accounting of 
    radiative transfer does not much matter; the narrative for
    super-puff formation told by \citet{paper3} should still hold.
    To create this figure, we resorted to the dust-free opacity table of
    \citet{freedman14} which extends to the cooler temperatures
    characterizing 1 AU at the cost of omitting heavy metals.
    For these opacities,
    $\gamma \equiv \kappa_{\rm vs}/\kappa_{\rm th} = 0.028 \ll 1$
    and so there is no stratospheric temperature inversion
    (cf.~Figure \ref{fig5}). Gas accretes at a somewhat slower 
    rate in the optically thin model than in the radiative diffusion
    model (note, as annotated, the difference in times
    at which the profiles are taken) 
    because the rcb of the former materializes at a deeper layer
    where densities and temperatures are higher,
    resulting in a lower cooling luminosity $L$. 
    These differences are probably overestimated by our model, which
    formally selects the innermost rcb to evaluate $L$;
    that innermost rcb is located, in the optically thin model, at the
    base of a tiny radiative window
    that opens up within the convective zone at $T \simeq 900$ K.
    We regard this feature as a possible artifact of our implementation
    of the \citet{piso14} method, in which it is not obvious which
    rcb to choose in evaluating $L$ when there are multiple rcb's.
    To obtain a more accurate $L$ (and one that may vary with position
    in the envelope) would require the use of a
    more sophisticated code such as \texttt{MESA} \citep{paxton11}.}
    \label{fig6}
\end{figure}

\section{Results}
\label{sec:result}

Figure \ref{fig4} reveals that the rate of gas accretion
does not much change when more careful
account is made of the outer optically thin
layer, as long as the outer nebular density $\rho_{\rm out}$ exceeds
$10^{-8} \rho_{\rm MMEN}$. 
For the same model parameters and at a given time, 
gas-to-core ratios in our new ``optically thin'' models
typically differ by less than a factor of 2 
from those of less realistic ``radiative diffusion'' models
(i.e., models that utilize the equation of radiative diffusion
even when the atmosphere is optically thin).
Cores can only accrete as much gas as can cool
(e.g., \citealt{paper2}), and the rate at which gas cools
is governed by the radiative-convective boundary (rcb).
That boundary, as demonstrated in Figure \ref{fig5}, 
sits well below the visible and thermal photospheres;
the temperature and pressure at the rcb remain practically unchanged
between most optically thin models and their radiative diffusion counterparts.
This is true whether the opacities are for dusty or dust-free gas.
Large temperature inversions appear in the uppermost layers,
effected by strong absorbers at visible wavelengths
(gas-phase atoms such as iron that render
$\gamma = \kappa_{\rm vs}/\kappa_{\rm th} \sim 240 \gg 1$; see section
\ref{ssec:opacity}). But this upstairs drama does not seem to much
affect the rcb downstairs. 

What sets the temperature and density of the rcb, and what connection
do they have, if any, to the outer optically thin layers?
The situation is easiest to understand for dusty envelopes,
whose rcb's are located where ${\rm H}_2$ starts to dissociate
\citep{paper1}.  Dissociation of H$_2$,
governed by Saha-type considerations,
mandates the rcb temperature to be $T_{\rm rcb} \simeq 2500\,{\rm K}$. 
As for the rcb density $\rho_{\rm rcb}$, that
is determined by 
both the rcb temperature and 
the interior adiabat.
In the convective zone below the rcb, the adiabatic index drops below
$4/3$ because energy is spent dissociating H$_2$ instead of heating the gas,
and the envelope mass
becomes concentrated toward the rocky core \citep{paper2}.  Thus 
$\rho_{\rm rcb}$ in dusty atmospheres is determined by the core radius,
the envelope mass, the internal adiabat, 
and $T_{\rm rcb}$,
none of which is
affected by the outer optically thin layers.

Dust-free envelopes are more complicated to diagnose.
To a crude approximation we can describe them as isothermal
(within a factor of $\sim$2) 
from the thermal photosphere down to the rcb 
(see bottom panels of Figure \ref{fig5}).
In the radiative diffusion models,
the temperature in this isothermal layer is set by the nebular temperature
$T_{\rm out} = 1000$ K.
For our new models with optically thin layers,
this temperature is set instead by the thermal photosphere:
$T_{\rm ph,th}^4 \sim T_{\rm int}^4 + T_{\rm eq}^4/2$, as follows
from inserting $\tau=2/3$ and $\gamma \gg 1$
into equation (\ref{eq5}). At 0.1 AU, $T_{\rm eq} \sim 883$ K
while $T_{\rm int} \sim 100$--200 K,
and so $T_{\rm ph,th} \sim 0.84\,T_{\rm eq} \sim 740$ K.
Since $T_{\rm ph,th}$ is close to $T_{\rm out}$,
it is not surprising that the new optically thin models behave
similarly to the old radiative diffusion models.
Technically, because of our temperature inversion,
$T_{\rm ph,th}$ is slightly lower than $T_{\rm out}$,
which in turn lowers $T_{\rm rcb}$ in the optically
thin models. The lower $T_{\rm rcb}$ and hence higher cooling luminosity
explain why the gas accretion rates in optically thin models 
are systematically higher than in the radiative diffusion
models; see how most of the solid curves in Figure \ref{fig4} are
higher than their dashed curve counterparts, not just in dust-free
models but also in dusty ones. Just the opposite effect occurs
when there is no temperature inversion, i.e., when $\gamma \ll 1$,
as in Figure \ref{fig6} discussed below.
We will explore further how $T_{\rm ph,th}$
tracks $T_{\rm out}$, and by extension how $T_{\rm rcb}$ 
is sensitive to the external
nebular environment for dust-free atmospheres,
in section \ref{ssec:1au}.

The situation changes once the ambient disc
is depleted in gas by more than eight orders of magnitude
relative to our fiducial gas-rich minimum-mass extrasolar nebula.
Under these extremely depleted conditions,
our treatment of the outer optically thin layers impacts more significantly
the growth and structure of the envelope as a whole.
In Figure \ref{fig4}, we see that for $\rho_{\rm out}/\rho_{\rm MMEN} = 10^{-8}$,
the final gas-to-core mass ratios
of the optically thin models (dusty or dust-free)
are approximately an order of magnitude
smaller than those of the corresponding radiative diffusion models. 
The reason for this drop is that the rcb has been pushed closer
to the surface of the underlying rock core, and with the envelope
more nearly completely radiative, differences between how we
treat the radiative transfer become magnified.
Extremely gas-poor environments
demand that atmospheric density profiles be sufficiently steep
to contain a given amount of gas within the planet's Hill sphere,
and this steepening pushes the thermal photosphere deeper into the envelope 
(compare the black dotted
curves in the right and second-from-right panels of Figure \ref{fig5});
the rcb is, of course, pushed inward as well.
For $\rho_{\rm out}/\rho_{\rm MMEN} = 10^{-8}$, the thermal photospheres
of the dusty and dust-free optically thin models
are only about twice the radius of the rocky core,
and the corresponding rcb densities and opacities are
2--10 times larger than those of the radiative diffusion models.
These larger rcb densities and opacities lower the radiative luminosities,
slowing cooling and by extension accretion---hence the especially low
gas-to-core ratios seen in Figure \ref{fig4}
characterizing $\rho_{\rm out}/\rho_{\rm MMEN} = 10^{-8}$. 
At even lower nebular densities (data not shown),
the rcb can be pushed all the way to the rocky core---i.e., the envelope
becomes purely radiative---at which point the model sequence terminates
(see also the isothermal endmember models in \citealt{paper2}, their Figure 4
and related discussion).

Although the models shown in Figure \ref{fig4} are technically
  those of nebulae whose gas densities are fixed from $t = 0$ to
  $t = 1$ Myr and drop to zero thereafter, they are still reliable
  guides for nebulae whose gas densities deplete smoothly over
  time. The very fact that rates of gas accretion change only by
  factors of $\sim$3 when the nebular density varies by 4 orders of
  magnitude suggests that the detailed time history of nebular
  depletion introduces, at most, only order-unity effects. This expectation is
  confirmed in Figure \ref{fig:gcr_v_t_evoldisk} which shows how
super-Earths amass $\sim$4--10\%-by-mass
gas envelopes in a nebula that decays exponentially with an
e-folding timescale of 0.1 Myr.

\section{Discussion}

\subsection{Creating Super-Puffs Beyond 1 AU}
\label{ssec:1au}

So far in this paper, we have concentrated exclusively on planets
located 0.1 AU from their central stars.
This restriction is due to our use of opacities
generated from the model of \citet{ferguson05}, which 
do not extend to the lower temperatures ($\lesssim 500$ K) 
characterizing more distant
regions of the disc. In this subsection we swap out the \citet{ferguson05}
opacities for the opacities from \citet{freedman14} which do not have this
limitation, in order to explore how atmospheric accretion
unfolds at larger stellocentric distances. The drawback is that 
the \citet{freedman14} opacity model
lacks the gas-phase heavy 
atomic metals (e.g., Fe)
that may dominate
the visible/ultraviolet wavelength opacity
(see section \ref{ssec:opacity}).\footnote{Whether such strong absorbers
in the visible are actually present is unclear. Introducing them
creates temperature inversions (see, e.g., our Figure \ref{fig5})
that might be present
in some extrasolar planets (e.g., the hot Jupiters WASP-33b and WASP-121b;
\citealt{haynes15}; \citealt{evans16}) but have not been
observed in others \citep[e.g., HD 209458b;][]{diamond-lowe14,line16}.
Heavy-metal atoms can gravitationally settle/rain out of atmospheres.
We have verified that our main result---that the cooling rates of optically thin models and radiative diffusion models differ by factors less than 2---is robust against the inclusion/exclusion of heavy metals.
}
We cannot expect the resultant calculation to be accurate,
but hope to identify rough, qualitative trends.

The specific question we want to address is whether, with our more
careful accounting of the outer optically thin layers,
the formation channel of ``super-puffs'' identified by \citet{paper3}
remains viable. Super-puffs are especially large ($R \gtrsim 4\,R_\oplus$) and low mass ($M \lesssim 6\,M_\oplus$) planets having gas-to-core ratios
of $\sim$20--40\%. In Lee \& Chiang (\citeyear{paper3}; see also
\citealt{inamdar15}), the envelopes of super-puffs are dust-free
and accrete at distances $\gtrsim 1$ AU. The requirement that the envelopes
be dust-free better couples the rcb temperature to the outer nebular
temperature (the buffering effects of dust are absent); that nebular
temperature, in turn, is lower at larger stellocentric
distances. 
Molecular ro-vibrational modes freeze out in colder gas so the opacity drops; more transparent envelopes cool faster and therefore grow faster, enabling
the production of super-puffs.
How does this picture change with a more realistic treatment
of the envelope's outermost, optically thin layers?

The answer is not by much,
as judged by Figure \ref{fig6}:
both our new optically thin model and the radiative diffusion
model are on their way to forming a super-puff planet
($M_{\rm core} = 2 M_\oplus$, GCR = 16\%). The models yield
similar interior structures and have similar gas accretion histories.

A key feature of our super-puff theory is the sensitivity of
the rcb to the external 
temperature (for dust-free atmospheres).
This sensitivity persists in our new optically thin treatment.
To within factors of 2, the rcb temperature is that of the deepest
photosphere. For $\gamma \gg 1$, the deepest photosphere
is the one evaluated at thermal (infrared) wavelengths: 
\begin{equation}
T_{\rm ph,th}^4 \sim T_{\rm int}^4 + T_{\rm eq}^4/2 \,.
\end{equation}
For $\gamma \ll 1$, the deepest photosphere is the
visible-light photosphere: 
\begin{equation}
T_{\rm ph,vs}^4 \sim T_{\rm int}^4/\gamma + T_{\rm eq}^4/\gamma
\end{equation}
obtained by substituting $\tau \sim 1/\gamma$ into equation (\ref{eq5});
recall that $\tau$ by definition is the optical depth at
thermal, not optical wavelengths.
In either case, as long as the external irradiation temperature
$T_{\rm eq}$ exceeds the internal temperature $T_{\rm int}$,
the rcb will respect
the former and cool down at larger stellocentric distances.

\begin{figure}
    \centering
    \includegraphics[width=0.5\textwidth]{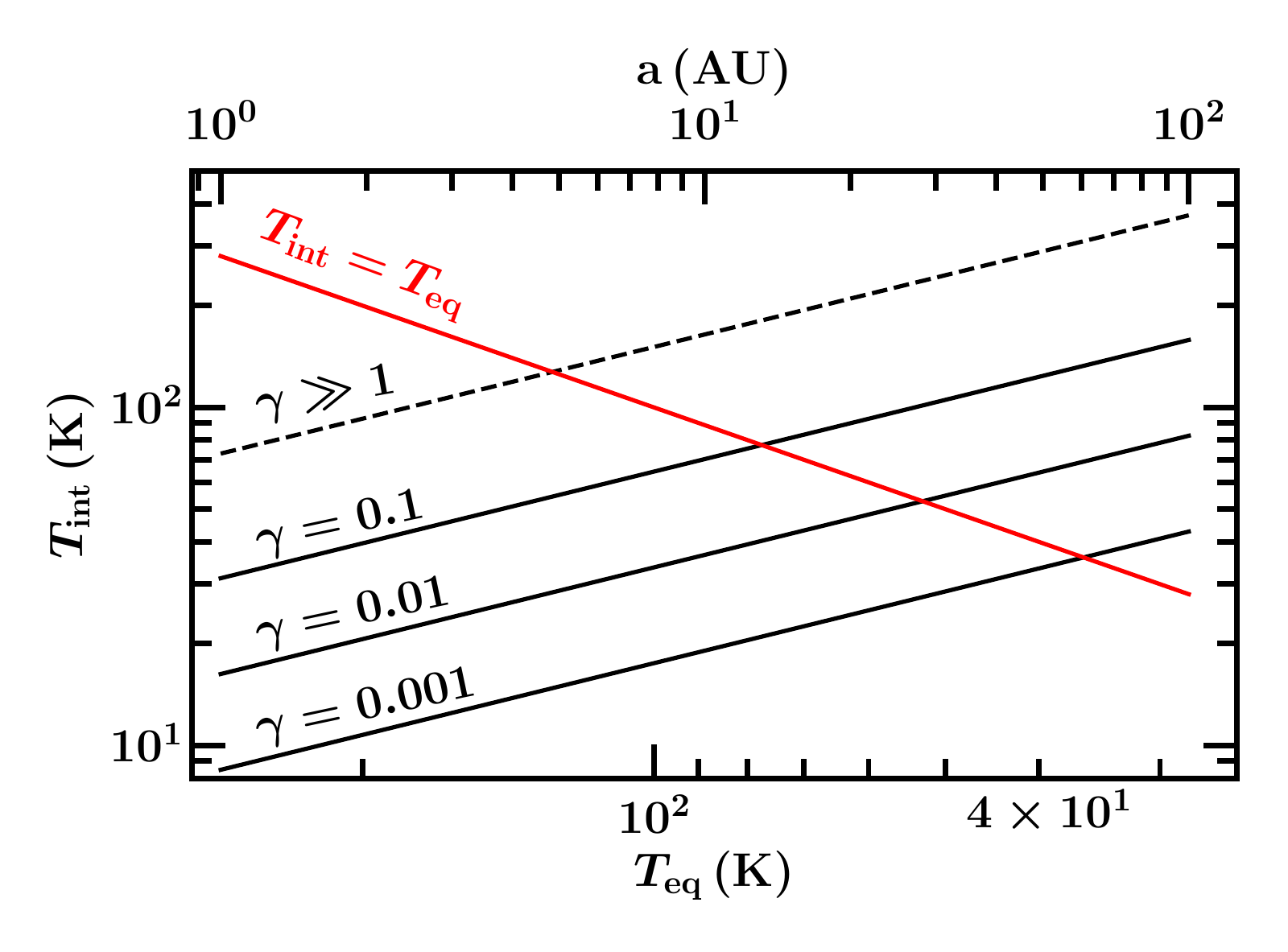}
    \caption{Estimating where in the disc we can expect
    the external irradiation temperature $T_{\rm eq}$ 
    to dominate the internal temperature $T_{\rm int}$
    in controlling the rcb temperature $T_{\rm rcb}$
    of nascent super-puffs ($M_{\rm core} = 2 M_{\oplus}$, GCR = 0.3).
    The black lines trace $T_{\rm int}$ vs.~$T_{\rm eq}$ for various
    values of $\gamma$ and are computed as follows: we first evaluate
    the cooling luminosity $L$ in equation (\ref{eq11}), taking
    $T_{\rm rcb} = T_{\rm eq}/2^{1/4}$ for $\gamma \gg 1$
    and $T_{\rm rcb} = T_{\rm eq} / \gamma^{1/4}$ for $\gamma < 1$;
    we then substitute this $L$,
    and $R_{\rm ph} = 0.9\,R_{\rm out}$ (where $R_{\rm out}$ equals
    the Bondi radius, which is smaller than the Hill radius for the parameters
    of this plot), into equation (\ref{eq6}) to solve
    for $T_{\rm int}$. This procedure assumes $T_{\rm eq} > T_{\rm int}$
    (so that $T_{\rm eq}$ controls $T_{\rm rcb}$) 
    and so the resultant black curves are only self-consistent up until
    they intersect the red $T_{\rm int} = T_{\rm eq}$ line.
    We see that the assumption $T_{\rm eq} > T_{\rm int}$ is valid within
    $\sim$5 AU for $\gamma \gg 1$ and out to larger distances for smaller
    $\gamma$. These are the distances out to which we can expect
    the rcb temperatures of dust-free planetary atmospheres to respect the
    external irradiation temperature, or in other words, the distances out
    to which we can expect the growth and evolution of atmospheres,
    including those of super-puffs, to be sensitive to the
    incident stellar flux.}
    \label{fig7}
\end{figure}

Far from the star, the inequality will reverse: the
irradiation temperature $T_{\rm eq}$
will fall below the internal temperature $T_{\rm int}$,
and the rcb may lose its sensitivity to the external environment.
We estimate at what stellocentric distance this happens by
assuming that $T_{\rm eq} > T_{\rm int}$ and seeing where this
assumption breaks down. The assumption implies that
\begin{equation} \label{relation1}
T_{\rm rcb} \sim T_{\rm ph,th} \sim T_{\rm eq}/2^{1/4}
\end{equation}
for $\gamma \gg 1$, and
\begin{equation} \label{relation2}
T_{\rm rcb} \sim T_{\rm ph,vs} \sim T_{\rm eq}/\gamma^{1/4}
\end{equation}
for $\gamma \ll 1$. To convert these into relations
between $T_{\rm eq}$ and $T_{\rm int}$, we replace
$T_{\rm rcb}$ with $T_{\rm int}$ using (\ref{eq6}),
in combination with 
equations (13) and (23) of \citet{paper2} for the
luminosity:
\begin{equation} 
    L \sim 4\times 10^{23}\,{\rm erg\,s^{-1}}\left(\frac{480\,{\rm K}}{T_{\rm rcb}}\right)^{4.5}\left(\frac{0.3}{{\rm GCR}}\right)^{1.6}\left(\frac{M_{\rm core}}{2\,M_\oplus}\right)^{4.6}
    \label{eq11}
\end{equation}
where we have taken 
the adiabatic index $\gamma_{\rm ad} = 1.3$.
Substituting (\ref{eq11}) into (\ref{eq6}) and
taking $R_{\rm ph} = 0.9\,R_{\rm out}$
(where $R_{\rm out}$ is the Bondi radius
for $M_{\rm core} = 2 M_\oplus$ and a gas temperature
of $T_{\rm out} = 1000 (a/0.1 \, {\rm AU})^{-3/7}$ K;
for $a=1$--100 AU, it is smaller than the Hill radius),
we solve for $T_{\rm int}$
in terms of $T_{\rm rcb}$, and from there obtain
relations between $T_{\rm int}$ and $T_{\rm eq}$
from (\ref{relation1}) and (\ref{relation2}). 
The resultant relationships between $T_{\rm int}$ and $T_{\rm eq}$ 
are shown in Figure \ref{fig7} as black lines.
The underlying assumption $T_{\rm eq} > T_{\rm int}$
is seen to be self-consistent out to $\sim$5 AU if $\gamma \gg 1$
and out to larger distances for smaller $\gamma$.
Inside these distances, the growth of super-Earth/super-puff
atmospheres respects the external nebular temperature $T_{\rm eq}$
(and not the nebular density);
outside these distances, we expect atmospheric growth to enter
an asymptotic regime that is insensitive to both nebular temperature
and nebular density.

\subsection{Atmospheric Mass Loss}
\label{ssec:accr_depl}

Our concern has been with how atmospheres grow during the gas disc
depletion phase. But atmospheres can also lose mass in
disc-depleted environments (\citealt{ikoma12, owen13, owen16, owen17,
  ginzburg16, ginzburg17}).  As we explain below, atmospheric erosion
generally takes place after, and does not interfere with, the 
atmospheric accretion phase that we have been exploring in this paper and earlier ones in our series.
 
In the process of cooling-limited accretion that we have been studying,
the cooling time of the atmosphere $t_{\rm cool}$ 
(equivalently, the gas accretion timescale)
adjusts to
the disc depletion timescale $t_{\rm disc}$. 
A necessary condition for the atmosphere to
erode instead of accrete is
$t_{\rm cool} > t_{\rm disc}$ \citep{ikoma12,owen16}.

Two ways have been discussed
in the literature to achieve this inequality.
In one way, $t_{\rm disc}$ suddenly becomes shorter
than $t_{\rm cool}$. In the ``two-timescale'' view of disc dispersal, the
nebula transitions from depleting on a slow timescale
to a fast one \citep{clarke01,alexander14}.
As soon as this switch occurs, the nebular
pressure that once confined the planetary envelope suddenly lifts,
allowing overpressured atmospheric gas to escape across the Hill/Bondi
boundary, either as a breeze or as a wind
(\citealt{ikoma12}; \citealt{owen16}). 
This way of triggering atmospheric mass loss
is not relevant for our scenario of late-time super-Earth
formation, which posits that rocky cores form after disc gas
has largely (but not completely) gone away. That is,
our story for super-Earth formation takes place wholly during the fast and
final second stage of nebular dispersal, 
after the switch has already occurred
(an explicit demonstration of atmospheric growth taking place entirely during the fast
  nebular dispersal phase is given by Figure
  \ref{fig:gcr_v_t_evoldisk}, which shows growth even when $t_{\rm
    disc}$ is as short as 0.1 Myr---the cooling timescale $t_{\rm
    cool}$ simply adjusts to match $t_{\rm disc}$).

The second way to achieve $t_{\rm cool} > t_{\rm disc}$
is by lengthening $t_{\rm cool}$. Long cooling times characterize
the latest atmospheric accretion stages,
when the nebula is most heavily depleted
and the radiative-convective boundary (rcb) of the envelope 
approaches the underlying rocky core 
(see our discussion toward the end 
of section \ref{sec:result} of the case 
$\rho_{\rm out}/\rho_{\rm MMEN} \leq 10^{-8}$).
When the inner convective zone shrinks to a thickness
comparable to the radius of the core,
it becomes possible for the envelope's thermal energy to fall below
the core's thermal energy---this is the case
when ${\rm GCR} \lesssim 5\%$ (\citealt{ginzburg16}; \citealt{ginzburg17}).
Under these conditions, it becomes energetically possible
for heat from the core to blow off the entire atmosphere (\citealt{ikoma12}). 
Whether it actually does so depends on a number of factors. First,
the planet must be close enough to its host star
that,
even though the rcb has retreated to near the core, 
the outer radiative envelope remains hot and distended out to the Hill/Bondi
radius, i.e., the wind's sonic point (embodied in equation 7 of \citealt{ginzburg17};
see also \citealt{fulton17} for possibly related observations).
Second, for the core to drive atmospheric mass loss effectively, it
must transport its energy
efficiently, by internal convection; a model assumption has been that the core
is isothermal at the same temperature as the gas right above it---that
the core and envelope cool in lockstep
at a rate controlled by the radiative-convective boundary of the
envelope (\citealt{ikoma12}; \citealt{ginzburg17}).  Whether core
convection might be too sluggish and fail to carry away enough heat
is an unresolved issue (see the large mantle viscosities considered by
\citealt{stamenkovic12} which imply core cooling timescales as long as
0.1--10 Gyr). A third factor is the efficiency with which 
core heat is converted into a mechanical wind; this efficiency has been
assumed to be unity (\citealt{ginzburg17}) but radiative losses from the
envelope will lower it.

If core power is not significant at driving mass loss, then
an outflow can still derive from the heat of the atmosphere itself
(\citealt{ikoma12}; \citealt{owen16}; \citealt{ginzburg16};
\citealt{ginzburg17}). Such winds are expected to be modest.
\citet{owen16} find that planetary gas mass fractions can be reduced 
by factors of 10 or more,
but these are overestimates insofar as most of the mass of their
convective atmospheres was assumed to be located at large radii.
More realistically, the adiabatic index $\gamma_{\rm ad} < 4/3$
in the convective interior,
which renders the atmospheric mass concentrated toward the core
(see section \ref{sec:introduction}). For reference, in most of our
accretionary models, $\lesssim 20\%$ of the envelope mass is in 
outer radiative zones.

Finally, the heating required to drive outflows may be provided
externally, by stellar photoionizing radiation at ultraviolet and
X-ray wavelengths which can heat atmospheric gas to temperatures
exceeding several thousand K.  Photoevaporative winds can shave away
or even destroy entirely the atmospheres of the closest-in
super-Earths, at orbital periods shorter than about 10 days
\citep{owen13,owen17, fulton17}.  Photoevaporative erosion takes place
on $\sim$100 Myr timescales, when stellar coronal activity remains
significant, and long after the entirety of the gas disc has been
purged (and replaced with a stellar wind).

\section{Summary}
\label{sec:conclusions}

Our study has shown that the theory of core accretion
can work just as well in optically thin environments
as it does in optically thick ones. Although core accretion
has traditionally been treated in wholly optically thick envelopes,
assumed to be embedded
in optically thick circumstellar discs (e.g., \citealt{pollack96}),
our calculations have demonstrated that how planetary atmospheres
grow does not much change when the
ambient disc density is reduced to values so low
that the disc and large portions of the envelope
are transparent. In particular,
super-Earths can accrete their $\sim$1\%-by-mass
atmospheres, and super-puffs their $\sim$20\%-by-mass
atmospheres, in discs depleted by many orders of magnitude---up to
seven in our model---relative to a conventional gas-rich
minimum-mass nebula. Such discs
can have vertical optical depths to visible-wavelength
starlight as low as $10^{-5}$ and radial optical
depths as low as $10^{-1}$ at stellocentric
distances of 0.1 AU.

\citet{paper3} have advocated the optically thin cavities
of ``transitional'' discs
as natural birth sites for super-Earths and their atmospheres.
Their calculations of core accretion were not 
always strictly self-consistent,
as they utilized the equation of radiative diffusion
to describe energy transport even when photons
in the thermal infrared were able to free stream
through gas envelopes.
Our work addressed this shortcoming by 
computing more realistically how radiation is transferred
in the outermost, optically thin layers of planetary atmospheres.
We found that such an accounting can have dramatic
impact on the envelope's outermost temperature structure, potentially
producing, e.g.,
strong stratospheric temperature inversions due to heavy
metals (e.g., atomic Fe). But such high-altitude complications
are more-or-less
immaterial for the cooling and hence gas accretion history of the planet.
Numerous studies (e.g., \citealt{piso14}; \citealt{paper2}; \citealt{paper3}; \citealt{ginzburg16}) have cited the primacy of the envelope's radiative-convective
boundary (rcb) in controlling that history. Our work reinforces
this point. We have taken care in the present work
to distinguish the visible and thermal
photospheres of the envelope, but the rcb,
lying at much greater depths,
behaves much as it does in earlier radiative diffusion models. In particular,
if gas is laden with dust, it remains true that the rcb properties
are determined by H$_2$ dissociation and not by any of the
details of the upper atmosphere; and if gas is free of dust, it is also still
true that the rcb temperature tends to follow the temperature of the upper
atmosphere, which is set by stellar irradiation for planets within a few AU
of their host stars.
Our work supports the picture of \citet{paper3}
that planets get their gas in parent discs that are themselves
nearly empty of gas; the rates of planetary gas accretion
that they computed are lower
than ours by factors on the order of 2. 
Only when there is so little nebular gas that the rcb is pushed all the way to the surface of the underlying rocky core 
(see also \citealt{ginzburg17}) 
does our gas accretion rate drop significantly compared to the previous work.

We have emphasized, here and in \citet{paper3}, how in dust-free atmospheres,
the rcb temperature tracks the temperature of the outer layers, specifically the temperature of the visible or infrared photosphere, whichever
is deeper. This connection implies that
more gas-rich planets---gas giants and super-puffs---form more easily
at large stellocentric distances where atmospheres are colder and therefore
optically thinner, cooling and growing more rapidly.
However, this connection between the nebular environment and the rcb eventually
breaks down. Far enough away from the host star, the incident radiation on the planet
will be too feeble to heat significantly the atmospheric layers above the rcb.
For these distant planets, the rcb temperature will asymptote to a value that
does not depend on stellocentric distance but will be set instead by the primordial
heat of formation (whatever energy was brought in by accreting gas). 
The factors controlling the ability of the disc to spawn gas-rich planets
will no longer include stellar irradiation, but will involve
the availability of solid materials and the efficiency with which
those solids can be assembled into cores. These are issues
that we are currently investigating.

\section*{Acknowledgments}

We are grateful to Richard Freedman and Mark Marley for in-depth conversations that helped us understand opacities better. We also thank Jeff Cuzzi, Kevin Heng, Jack Lissauer, Nikku Madhusudhan, Eliot Quataert, and Leslie Rogers 
for helpful and motivating discussions. 
Andrew Youdin provided positive and constructive feedback on our submitted
manuscript, and Michiel Lambrechts delivered a thoughtful and encouraging
referee's report that led to substantive improvements. 
EJL was supported in part by NSERC of Canada under PGS D3, the Berkeley Fellowship, and the Sherman Fairchild Fellowship at Caltech. EC acknowledges support from the NSF.
This research used the Savio computational cluster resource provided by the Berkeley Research Computing program at the University of California, Berkeley, supported by the UC Berkeley Chancellor, Vice Chancellor for Research, and Chief Information Officer.

\bibliographystyle{mnras}
\bibliography{optthin}

\label{lastpage}

\end{document}